\documentclass[a4paper,11pt]{article}

\usepackage{amsthm,amsmath,amsfonts,amssymb,xspace}

\theoremstyle{definition}
\newtheorem{definition}{Definition}
\newtheorem{example}[definition]{Example}
\theoremstyle{plain}
\newtheorem{theorem}[definition]{Theorem}
\newtheorem{lemma}[definition]{Lemma}
\newtheorem{proposition}[definition]{Proposition}

\renewcommand{\geq}{\geqslant}
\renewcommand{\leq}{\leqslant}

\newcommand{\abar}{\bar{a}}
\newcommand{\vbar}{\bar{v}}
\newcommand{\xbar}{\bar{x}}

\newcommand{\calD}{\mathcal{D}}

\newcommand{\reducesto}{\leq_{\mathrm{T}}}
\newcommand{\equivto}{\equiv_{\mathrm{T}}}

\newcommand{\GFtwo}{\ensuremath{\mathrm{GF}(2)}\xspace}

\newcommand{\polyk}[1][k]{\mathcal{P}_{#1}}
\newcommand{\ratfunc}[1][]{\mathcal{F}_{\!\mathrm{B}}^{#1}}

\newcommand{\rationals}{\mathbb{Q}}
\newcommand{\ints}{\mathbb{Z}}

\newcommand{\Pclass}{\ensuremath{\mathsf{P}}\xspace}
\newcommand{\NP}{\ensuremath{\mathsf{NP}}\xspace}
\newcommand{\FP}{\ensuremath{\mathsf{FP}}\xspace}
\newcommand{\numP}{\ensuremath{\mathsf{\#P}}\xspace}
\newcommand{\numPQ}{\ensuremath{\numP_{\rationals}}\xspace}
\newcommand{\FPnumP}{\ensuremath{\FP^{\numP}}\xspace}

\newcommand{\CSP}{\ensuremath{\mathrm{CSP}}\xspace}
\newcommand{\CSPclass}{\ensuremath{\mathsf{CSP}}\xspace}
\newcommand{\numCSP}{\ensuremath{\mathrm{\#CSP}}\xspace}
\newcommand{\numCSPclass}{\ensuremath{\mathsf{\#CSP}}\xspace}
\newcommand{\numBIS}{\ensuremath{\mathrm{\#BIS}}\xspace}
\newcommand{\numSAT}{\ensuremath{\mathrm{\#SAT}}\xspace}

\newcommand{\negquad}{\!\!\!\!}
\newcommand{\Eval}[1]{\textsc{Eval}(#1)}
\newcommand{\transp}[1]{#1^\mathrm{T}}

\newcommand{\smat}[1]{\ensuremath{\bigl(\begin{smallmatrix}#1\end{smallmatrix}\bigr)}}

\title{The Complexity of Weighted Boolean\\
        \#CSP with Mixed Signs\thanks{~Funded in part by the EPSRC grant ``The Complexity of Counting in Constraint Satisfaction Problems''.}
}

\author{Andrei Bulatov,%
        \thanks{~\texttt{abulatov@cs.sfu.edu}\,.  School of Computing
                 Science, Simon Fraser University, Burnaby, Canada.}
      \ Martin Dyer,%
        \thanks{~\{\texttt{dyer,richerby}\}\texttt{@comp.leeds.ac.uk}\,.
                 School of Computing, University of Leeds, Leeds, LS2~9JT, UK.}
      \ Leslie Ann Goldberg,%
        \thanks{~\{\texttt{L.A.Goldberg,M.Jalsenius}\}%
                 \texttt{@liverpool.ac.uk}\,.  Department of Computer
                 Science, University of Liverpool, Liverpool, L69~3BX, UK.}
       \\Markus Jalsenius\footnotemark[4]
       \ and David Richerby\footnotemark[3]                 
}

\begin{document}

\maketitle

\begin{abstract}
    We give a complexity dichotomy for the problem of computing the
    partition function of a weighted Boolean constraint satisfaction
    problem.  Such a problem is parameterized by a set $\Gamma$ of
    rational-valued functions, which generalize constraints.  Each
    function assigns a weight to every assignment to a set of Boolean
    variables.  Our dichotomy extends previous work in which
    the weight functions were restricted to being non-negative.  We
    represent a weight function as a product of the form $(-1)^s g$,
    where the polynomial $s$ determines the sign of the weight and the
    non-negative function $g$ determines its magnitude.  We show that
    the problem of computing the partition function (the sum of the
    weights of all possible variable assignments) is in
    polynomial time if either every function in $\Gamma$ can be
    defined by a ``pure affine'' magnitude with a quadratic sign
    polynomial or every function can be defined by a magnitude of
    ``product type'' with a linear sign polynomial.  In all other
    cases, computing the partition function is \FPnumP-complete.
\end{abstract}


\section{Introduction}

The principal result of this paper is a dichotomy theorem for the
complexity of computing the partition function of a weighted Boolean
constraint satisfaction problem. This problem has a set of functions
$\Gamma$ that are used to assign a weight to any
configuration, where a configuration is an assignment of values to the
instance's variables. These functions generalize
constraint relations in the classical constraint satisfaction problem
(CSP), which corresponds to the case where all functions in $\Gamma$ have
range $\{0, 1\}$. The problem we consider here is to compute the
\emph{partition function} of a given instance of weighted CSP;
that is, the sum of weights of all configurations. Computing the
partition function generalizes the problem of counting the number of
satisfying solutions of a CSP. We denote by $\numCSP(\Gamma)$ the
problem of computing the partition function of weighted CSP instances
which use functions from the set $\Gamma$.

The term ``partition function'' originates in statistical physics, and
certain problems from statistical physics may be expressed as weighted
CSPs. For example, the Potts model~\cite{Welsh93:knots} can be
expressed as a weighted CSP, whereas only the ``hard core'' version can
be expressed as a classical CSP. The two possible hard core versions
of the Potts model correspond to graph colouring, in the
so-called antiferromagnetic case, and the trivial problem of colouring
each component of a graph with a single colour, in the
so-called ferromagnetic case.

Here, we extend the work of Dyer, Goldberg and
Jerrum~\cite{DGJ:weighted-Boolean}, who gave a dichotomy for
the complexity of $\numCSP(\Gamma)$ when every function in $\Gamma$ is
restricted to have non-negative values. They defined two classes of
functions, those that are ``pure affine'' and those of ``product
type'', and showed that $\numCSP(\Gamma)$ is in $\FP$ if, and only if,
every function in $\Gamma$ is pure affine or every function is of product
type. Otherwise $\numCSP(\Gamma)$ is complete for $\FPnumP\!$. The
existence of algorithms for testing the properties of being purely
affine or of product type means that the dichotomy is
decidable.

The contribution of this paper is a dichotomy theorem for
$\numCSP(\Gamma)$, where $\Gamma$ is allowed to contain function which
give values of either sign.  Specifically, $\numCSP(\Gamma)$
is either in $\FP$ or is $\FPnumP$-complete.  As in the
non-negative case, the dichotomy is decidable.

This extension is of particular interest
because functions having mixed signs can cause cancellations in the
partition function, which may make it easier to compute.  Many
natural problems can be expressed as weighted $\numCSPclass$ problems
with functions of mixed signs.  For example, if $f$ is a binary
function, an instance $I$ of $\numCSP(\{f\})$ corresponds to a
graph $G_I$ where each variable of $I$ is a vertex and each constraint
corresponds to an edge.  There is a binary function $f\colon
\{0,1\}^2\to \{-1, 1\}$ such that the partition function of
$\numCSP(\{f\})$ counts the number of subgraphs in $G_I$ that have an
even number of edges --- see the examples in
Section~\ref{sec:Intro:Related} for details.


\subsection{Constraint satisfaction}

\emph{Constraint satisfaction} provides a general framework for
modelling decision problems and has many practical applications,
particularly in artificial intelligence --- see, for example,
\cite{RBW06:constraint-programming}.  Decisions are modelled by
variables, which are subject to constraints that model the logical and
resource restrictions. Many interesting problems can be modelled in
this way, including problems in the areas of satisfiability,
scheduling and graph-theory. Consequently, the computational
complexity of constraint satisfaction problems has become a major and
active area of research~\cite{CKS01:boolean-csp,
HN04:graph-homomorph}.

A constraint satisfaction problem (CSP) has a finite \emph{domain},
which we may denote by $\{0, 1, \dots, q - 1\}$ for some positive
integer $q$. In this paper we are interested only in the
\emph{Boolean} case, where $q = 2$. A \emph{constraint language}
$\Gamma$ with domain $\{0, 1, \dots, q - 1\}$ is a set of relations on
$\{0, 1, \dots, q - 1\}$. For example, let $q = 2$, and consider the
relation $R = \{(1, 0, 0), (1, 0, 1), (0, 1, 0), (0, 1, 1)\}$. This is
a 3-ary relation on the domain $\{0, 1\}$, having four tuples.

Given a constraint language $\Gamma$, an \emph{instance} of
$\CSP(\Gamma)$ is a set of \emph{variables} $V = \{v_1, \dots, v_n\}$
and a set of \emph{constraints}. Each constraint has a \emph{scope},
which is a tuple of variables and a
relation from $\Gamma$ of the same arity, which constrains the
variables in the scope.  A \emph{configuration} $\sigma$ is a function
from $V$ to $\{0, 1, \dots, q - 1\}$. The configuration $\sigma$ is
\emph{satisfying} if the scope of every constraint is mapped to a
tuple that is in the corresponding relation. In our example above, a
configuration $\sigma$ satisfies the constraint with scope $(v_3, v_7,
v_2)$ and relation $R$ if, and only if, it maps exactly one of $v_3$
and $v_7$ to the value~1.  For a CSP with constraint language
$\Gamma$, the decision problem $\CSP(\Gamma)$ is to determine whether
a given instance $I$ has a satisfying configuration. The counting
problem $\numCSP(\Gamma)$ is to determine the number of distinct
satisfying configurations of $I$.

Varying the constraint language $\Gamma$ defines the classes
$\CSPclass$ and $\numCSPclass$ of decision and counting
problems. These contain problems of very different computational
complexity. For example, if $\Gamma = \{ R_1 , R_2 , R_3 \}$ where
$R_1$, $R_2$ and $R_3$ are the three binary relations defined by $R_1
= \{(0, 1), (1, 0), (1, 1)\}$, $R_2 = \{(0, 0), (0, 1), (1, 1)\}$ and
$R_3 = \{(0, 0), (0, 1), (1, 0)\}$, then $\CSP(\Gamma)$ is the
classical 2-Satisfiability problem, which is in $\Pclass$. On the
other hand, there is a similar constraint language that expresses
3-Satisfiability, which is $\NP$-complete. There are cases where the
counting problem is harder
than the decision problem: if $\Gamma$ is the constraint language
defining 2-Satisfiability, then $\numCSP(\Gamma)$ contains the problem
of counting independent sets in graphs, which is
$\numP$-complete~\cite{Val79:enumeration}, even for 3-regular
graphs~\cite{Green00:counting-sparse}.

Any problem in $\CSPclass$ is in $\NP$, but not every problem in $\NP$
can be expressed in $\CSPclass$. For example, the question ``Is the
graph $G$ Hamiltonian?''\@ cannot be expressed in $\CSPclass$, because
the property of being Hamiltonian cannot be captured by constraints of
fixed size. This is a limitation of the class $\CSPclass$,
but it also
has an advantage. If $\Pclass \neq \NP$, there are problems which are
neither in $\Pclass$ nor
$\NP$-complete~\cite{Lad75:reducibility} but, for smaller classes of
decision problems, the situation may be more straightforward. A
dichotomy theorem may be possible, partitioning all problems in the
class into those which are in $\Pclass$ and those which are
$\NP$-complete, with no problems of intermediate complexity. It has
been conjectured, in the seminal paper of Feder and
Vardi~\cite{FV99:datalog}, that there is a dichotomy theorem for
$\CSPclass$. Although much progress has been made towards proving
this, it remains unproven to date.

In the Boolean case, the status of $\CSPclass$ was resolved by
Schaefer~\cite{Schaefer78:satisfiability}. Schaefer proved a dichotomy
for the domain $\{0, 1\}$, giving four conditions on the constraint
language $\Gamma$. If any of the conditions holds then $\CSP(\Gamma)$ is
in $\Pclass$, otherwise $\CSP(\Gamma)$ is $\NP$-complete. For details,
the interested reader is referred to Schaefer's
paper~\cite{Schaefer78:satisfiability} or to Theorem~6.2 of the
textbook~\cite{CKS01:boolean-csp}. An interesting feature is that
Schaefer's conditions are all algorithmically checkable. Thus, given a
constraint language $\Gamma$ with domain $\{0, 1\}$, we can determine
whether $\CSP(\Gamma)$ is in $\Pclass$ or $\NP$-complete.

While the conjectured dichotomy for $\CSPclass$ remains open,
Bulatov~\cite{Bul08:counting-csp} has recently made a major
breakthrough for $\numCSPclass$. He has shown that there is a
dichotomy between $\FP$ and $\numP$-complete, for the whole of
$\numCSPclass$. However, his proof sheds
very little light on when $\numCSP(\Gamma)$ is in $\FP$, and when it is
$\numP$-complete. The difficulty is that, while $\Gamma$ itself is of
fixed size, the criterion of the dichotomy involves
finding a defect in any of a potentially infinite class of structures
built on $\Gamma$. Whether this criterion is algorithmically checkable
is an open question.

In the Boolean case, which is our focus here, a decidable dichotomy
theorem for $\numCSPclass$ had already been established by Creignou
and Hermann~\cite{CH96:counting-problems}.  Before stating their
theorem we introduce the following definition. A Boolean relation $R$
is \emph{affine} if it is the set of solutions to a system of linear
equations over $\GFtwo$. A constraint language $\Gamma$ is affine if
every relation $R \in \Gamma$ is affine. Creignou and Hermann prove
that $\numCSP(\Gamma)$ is in $\FP$ if $\Gamma$ is affine and is
$\numP$-complete, otherwise. There is an algorithm that determines
whether a
Boolean constraint language $\Gamma$ is affine, so there is an
algorithm that determines whether $\numCSP(\Gamma)$ is in $\FP$ or
$\numP$-complete. In addition to Creignou and Hermann's dichotomy,
Dyer, Goldberg and Jerrum~\cite{DGJ:trichotomy-boolean} have given an
approximation trichotomy for Boolean $\numCSPclass$. Let $\numBIS$
denote the problem of counting the number of independent sets in a
bipartite graph and let $\numSAT$ denote the problem of counting
satisfying assignments to a Boolean formula in conjunctive normal
form. Dyer, et al.\@~\cite{DGJ:trichotomy-boolean} have shown that if
$\Gamma$ is not affine (hence $\numCSP(\Gamma)$ is $\numP$-complete)
then there is an approximation-preserving reduction between
$\numCSP(\Gamma)$ and either $\numBIS$ or $\numSAT$.


\subsection{Weighted $\numCSPclass$}

The counting problem $\numCSP(\Gamma)$ can be extended naturally by
replacing the relations in $\Gamma$ by functions. We refer to
the corresponding class of problems as \emph{weighted
$\numCSPclass$}. The functions are used to assign weights to
configurations and the \emph{partition function} computes the sum of the
weights over all configurations. We give a formal definition
below. The partition function of a weighted $\numCSP$ generalizes
the number of satisfying solutions of a classical $\numCSP$. The
classical setting may be recovered by restricting the range of
every function to $\{0,1\}$.

In weighted $\numCSP$, a \emph{constraint language} over a finite
domain $\calD$ is a finite collection of functions $\Gamma =
\{f_i\colon \calD^{r_i}\to \rationals \mid i\in I\}$.  The
natural number $r_i$ is called the \emph{arity} of the function $f_i$;
we refer to functions of arity one, two and three as unary,
binary and ternary, respectively. In this paper we
consider exclusively the Boolean domain, $\calD = \{0, 1\}$.

An \emph{instance} of a weighted constraint satisfaction problem over
a constraint language $\Gamma$ is a pair $I = (V, C)$, where $V =
\{v_1, \dots, v_n\}$ is a set of \emph{variables} and $C$ is a finite
set of \emph{constraints}.  Each constraint is of the form $f(v_{i_1},
\dots, v_{i_r})$, where $f$ is an $r$-ary function in the set
$\Gamma$.  To keep notation simple, we will often use $x_1, x_2,
\dots$ as ``metavariables'', standing for variables in $V\!$.

A \emph{configuration} of an instance $(V, C)$ is a function
$\sigma\colon V\to \calD$, assigning a value from the domain to each
variable.  The \emph{weight} of a configuration~$\sigma$ is defined to be
\begin{equation*}
    W(\sigma) := \negquad\negquad\!\!
                 \prod_{f(x_1, \dots, x_r)\in C}
                 \negquad\negquad\!\!
                                  f(\sigma(x_1), \dots, \sigma(x_r))\,.
\end{equation*}

We are interested in computing the \emph{partition function} of an
instance $I$.  This is the sum $Z(I)$ of the weights of all possible
configurations:
\begin{equation*}
    Z(I) := \negquad \sum_{\sigma\colon V\to \calD}\negquad W(\sigma)\,.
\end{equation*}
The \emph{weighted constraint satisfaction problem} is the problem of
computing $Z(I)$ given an instance $I$. Since this paper is
exclusively about weighted $\CSP$s, we will drop the word ``weighted'' and
write $\numCSP(\Gamma)$ for the weighted constraint satisfaction
problem over the constraint language $\Gamma$ and $\numCSPclass$ for
the union of $\numCSP(\Gamma)$ over all rational-weighted constraint
languages.  For constraint languages with only a single function $f$,
we write $\numCSP(f)$, rather than $\numCSP(\{f\})$.

A constraint $f(x_1, \dots, x_r)$ is \emph{satisfied} by a
configuration $\sigma$ if $f(\sigma(x_1), \dots,\nolinebreak[3]
\sigma(x_r))\neq 0$. Therefore, the weight of a configuration is zero
unless it satisfies every constraint.  If we restrict to constraint
languages where every function has range $\{0, 1\}$, the weight of
every configuration is either zero or one and $Z(I)$ is just the
number of satisfying configurations for $I$.  This corresponds
precisely to the counting constraint satisfaction problem.


\subsection{Related work}
\label{sec:Intro:Related}

Bulatov's counting dichotomy~\cite{Bul08:counting-csp} can be extended
to weighted $\numCSPclass$ as long as the range of every function $f
\in \Gamma$ is $\rationals^{\geq 0}$ (the set of non-negative
rationals)~\cite{BDGJ:personal-communication}. However, it is not
known whether it extends to weighted $\numCSPclass$ with functions of
mixed signs. Furthermore, there is currently no algorithm known that
determines whether $\numCSP(\Gamma)$ is in $\FP$ or
$\FPnumP$-complete, given a constraint language $\Gamma$.  For the
special case of graph homomorphisms, an effective dichotomy is
known for functions of mixed signs~\cite{GGJT:mixed-signs}. There is
also a dichotomy theorem by Dyer et al.\@ for Boolean weighted
$\numCSPclass$ for functions that are
non-negative~\cite{DGJ:weighted-Boolean}. This is expressed in terms
of two classes of functions, \emph{pure affine} and \emph{product
type}, which we define in
Section~\ref{sec:classes-of-functions}. Dyer et
al.\@ give the following theorem.

\begin{theorem}[{\cite[Theorem~4]{DGJ:weighted-Boolean}}]
\label{thm:non-negative-theorem}
    Let $\Gamma$ be a constraint language in which the range of
    every function $f \in \Gamma$ is a set of non-negative
    rationals. If every function in $\Gamma$ is pure affine, then
    $\numCSP(\Gamma)\in \FP$. If every function in $\Gamma$ is of
    product type, then $\numCSP(\Gamma)\in \FP$. Otherwise,
    $\numCSP(\Gamma)$ is $\FPnumP$-complete.
\end{theorem}

There exist algorithms that test whether a Boolean constraint
language $\Gamma$ is pure affine or of product type. This means that
the dichotomy is effectively decidable.

The contribution of this paper (Theorem~\ref{thm:main-theorem} below)
extends Theorem~\ref{thm:non-negative-theorem} to constraint languages
$\Gamma$ containing arbitrary rational-valued functions. This is an
interesting extension since functions with negative values can cause
cancellations and may make the partition function easier to
compute.\footnote{In a related context, recall the sharp distinction in
complexity between computing the permanent and the determinant of
a matrix.}  Independently, Cai, Lu and Xia have recently found a wider
generalization, giving a dichotomy for the case where $\Gamma$ can be
any set of complex-valued functions~\cite{CLX:complex}.

The case of mixed signs has been been considered previously by
Goldberg, Grohe, Jerrum and Thurley~\cite{GGJT:mixed-signs}, in the
case of one symmetric binary function on an arbitrary finite
domain. Their theorem generalizes that of Bulatov and
Grohe~\cite{BG2005:partition} for the non-negative case. Goldberg et
al.~\cite{GGJT:mixed-signs} give two examples, which can also be
expressed as Boolean weighted $\numCSPclass$, and fall within the
scope of this paper. The first appeared as an open problem
in~\cite{BG2005:partition}. The complexity of these problems can be
deduced from~\cite{GGJT:mixed-signs} and from the results of this
paper.

\begin{example}
\label{example:xy}
    The first example in~\cite{GGJT:mixed-signs} is the function
    $f\colon \{0,1\}\to \{-1,1\}$, where
    \begin{align*}
        f(0,0) &= 1 & f(0,1) = \phantom{-}1\phantom{\,.} \\
        f(1,0) &= 1 & f(1,1) = -1\,.
    \end{align*}
    An instance $I$ of $\numCSP(f)$ can be represented by a graph $G =
    (V,E)$ with $n$ vertices.  (In fact the argument remains the same
    even in the case where $G$ is a multigraph with self-loops.)  The
    set of variables in $I$ is $V$ and, for each edge
    $(u,v) \in E$, we have the constraint $f(u,v)$ in $I$. Then
    $\frac{1}{2}Z(G) + 2^{n-1}$ is the number of induced subgraphs of
    $G$ with an even number of edges. Hence, up
    to a simple transformation, the partition function $Z(G)$ counts
    induced subgraphs with an even number of edges. To see this,
    observe that for every configuration $\sigma$, the term
    $\prod_{(u,v) \in E} f(u,v)$ is 1 if the subgraph of $G$ induced
    by $\sigma^{-1}(1)$ has an even number of edges and $-1$
    otherwise. In terms of our Theorem~\ref{thm:main-theorem} below,
    $f(x,y)=(-1)^{xy}$, so this problem is in $\FP$, and an algorithm
    for computing $Z(G)$ follows from Lemma~\ref{lemma:easy-affine}
    below.
\end{example}

\begin{example}
\label{example:x+y}
    The second example in~\cite{GGJT:mixed-signs} is $\numCSP(f)$,
    where
    \begin{align*}
        f(0,0) &= \phantom{-}1 & f(0,1) =           -1\phantom{\,.} \\
        f(1,0) &=           -1 & f(1,1) = \phantom{-}1\,.
    \end{align*}
    In terms of Theorem~\ref{thm:main-theorem} below,
    $f(x,y)=(-1)^{x+y}$, so this problem is also in $\FP$. This can
    easily be shown directly. Let $G = (V,E)$ be a graph with $n$
    vertices. Note that $Z(G)$ is unchanged by removing any circuit
    from $G$. Thus we may reduce $G$ to a forest $F$, which will have
    edges if, and only if, $G$ was not Eulerian. If $F$ has no edges,
    then $Z(G)=Z(F)=2^n$. Otherwise $F$ has at least one leaf vertex
    $v$, and then we have $Z(G)=Z(F)=Z(F\setminus v)-Z(F\setminus
    v)=0$.  Thus $Z(G)=0$ unless $G$ is Eulerian, in which case
    $Z(G)=2^n$, and hence the problem is trivially in \FP.
\end{example}


\subsection{Complexity}

Since our weights are arbitrary rationals, the partition function $Z$
is not, in general, an integer-valued function.  As such, $\numCSP(\Gamma)$
is not, in general, in the class \numP.  However, it is easy to see
that, for every constraint language $\Gamma$, there is a partition
function $Z'$ in \numP and an \FP-computable integer-valued function $K$
such that, for all instances $I$ of $\numCSP(\Gamma)$,
$Z(I) = Z'(I)/K(I)$.  This is achieved by ``clearing denominators''
(see~\cite{GJ:Tutte-inapprox}).

Following \cite{GJ:Tutte-inapprox}, we write $\numPQ$ for the class of
functions of the form $f/g$, where $f\in \numP$ and $g\in \FP$.  It is
immediate that
\begin{equation*}
    \numCSPclass \subseteq \numPQ \subseteq \FPnumP.
\end{equation*}

\begin{proposition}
    Every \numCSPclass problem that is \numP-hard is
    \FPnumP-complete.
\end{proposition}
\begin{proof}
    If $Z(I)\in \numCSPclass$ is \numP-hard, we can use an oracle for
    $Z(I)$ to construct an oracle for $Z'(I)$, as described above.
    With this oracle, we can compute any problem in \FPnumP in a
    polynomial number of steps.  Therefore, $Z(I)$ is
    \FPnumP-complete.
\end{proof}

Let $\Gamma$ be a constraint language.  We say that $\Gamma$
\emph{simulates} a function $f\notin\Gamma$ if, given an instance $I$
of $\numCSP(\Gamma \cup \{f\})$, we can construct, in polynomial time,
an instance $I'$ of $\numCSP(\Gamma)$ such that $Z(I) = K(I)Z(I')$
for some \FP-computable function $K$.  This generalizes
parsimonious reductions \cite{Pap1994:Complexity}; clearly, if
$\Gamma$ simulates $f$ then $\numCSP(\Gamma \cup \{f\}) \reducesto
\numCSP(\Gamma)$, where $\reducesto$ denotes polynomial-time Turing
reducibility.  We write $\numCSP(\Gamma)
\equivto \numCSP(\Gamma')$ in the case that  $\numCSP(\Gamma)\reducesto
\numCSP(\Gamma')\reducesto \numCSP(\Gamma)$.


\subsection{Organization of the paper}

Our paper is organized as follows.  In Section~\ref{sec:Background},
we define notation and the classes of functions we will use
throughout the paper.  In Section~\ref{sec:Dichotomy}, we state our
dichotomy result and prove the polynomial-time cases.  The remaining
sections prove that all other cases are \FPnumP-complete.  We give
useful tools for proving hardness in Sections~\ref{sec:Reductions}.
In Sections~\ref{sec:Hard-affine} and~\ref{sec:Hard-product}, we show,
respectively, that any constraint language containing a pure affine
function of degree greater than~2 is \numP-hard and that and any
language with a function of product type of degree greater than~1 can
be made \numP-hard by adding a simple function.  Finally, we
complete the proof of the dichotomy in Section~\ref{sec:Proof},
showing that the simple function is can be simulated by the
functions already present.


\section{Some notation}
\label{sec:Background}

All sets and other objects referred to in this paper are finite unless
it is stated otherwise.  We write $\abar$ for a tuple of elements
$(a_1, \dots, a_r)$ for some $r$ and, for natural numbers $m\leq n$,
we write $[m,n]$ for the set $\{m, m+1, \dots, n\}$.

The \emph{support} of a function $f\colon X^r \to \rationals$ is the
$r$-ary relation $\{\abar \mid f(\abar)\neq 0\}$.  For a function
$g\colon X\rightarrow Y$ and a tuple $\abar\in X^r\!$, we write
$g(\abar)$ for the tuple $(g(a_1), \dots, g(a_r))$.

We write $\ratfunc$ for the set of all functions, of all positive
arities, from the set $\{0, 1\}$ to $\rationals$, the rationals, and
$\ratfunc[\geq0]$ for the subset of $\ratfunc$ consisting of all
functions with non-negative ranges.  We write $\polyk$ for the set of
multivariate polynomials in variables $x_1, \dots, x_k$ over $\GFtwo$.
We sometimes write $p(x_1, \dots, x_k)$ for a polynomial $p\in\polyk$
or other function, to emphasize that $p$ is a function of those
variables.

A function $f(x_1, \dots, x_k)$ \emph{depends on} a variable $x_i$ if there are constants
$c_1, \dots, c_{i-1}, \nolinebreak[3] c_{i+1}, \dots, c_k\in \{0, 1\}$ such that
\begin{equation*}
    f(c_1, \dots, c_{i-1}, 0, c_{i+1}, \dots, c_k)
                 \neq f(c_1, \dots, c_{i-1}, 1, c_{i+1}, \dots, c_k)\,.
\end{equation*}


\subsection{Classes of functions}
\label{sec:classes-of-functions}

In this section, we define the classes of functions that we use
throughout the paper.  Our definitions of pure affine functions and
functions of product type are those used by Dyer et
al.\@ \cite{DGJ:weighted-Boolean} but multiplied by a term $(-1)^s$ for
some polynomial $s$, which determines the sign.

Recall that a relation over $\{0, 1\}$ is \emph{affine} if it is the
solution set of a set of linear equations over $\GFtwo$.  We say that
a function $f\in \ratfunc$ is affine if it has affine support.

\begin{definition}
\label{defn:pure-affine}
    A $k$-ary function $f\in\ratfunc$ is \emph{pure affine} if there
    is a constant $w\in \rationals^{>0}\!$, an affine function $g\in
    \polyk$ and a polynomial $s\in \polyk$ such that
    \begin{equation}
    \label{eq:pure-affine}
        f(\xbar)\ =\
            w\hspace{1pt}(-1)^{s(\xbar)} g(\xbar)\,.
    \end{equation}
\end{definition}

Note that the range of $f$ in (\ref{eq:pure-affine}) is included in
$\{-w, 0, w\}$.  The polynomial $g$ is uniquely defined, up to the
identities $x\oplus x = 0$ and $x^2=x$.  However, because the value of
$f$ does not depend on the value of $s$ for values of its inputs where
$g(\xbar) = 0$, there may be several distinct polynomials $s$ for
which the identity (\ref{eq:pure-affine}) holds.  If $s$ is of minimal
degree $d$ such that (\ref{eq:pure-affine}) holds, we say that $s$ is
\emph{degree-minimized} with respect to $g$ and that $f$ is \emph{pure
    affine of degree $d$}.  For the purposes of this paper, we
consider the constant zero and one polynomials to have degree zero.

We write $\chi_{=}$ and $\chi_{\neq}$ for the binary equality and
disequality functions, respectively, defined as
\begin{equation*}
    \chi_{=}(x,y) = x\oplus y\oplus 1 \quad\quad\quad\quad
    \chi_{\neq}(x,y) = x\oplus y\,.
\end{equation*}

\begin{definition}
\label{defn:product-type}
    A $k$-ary function $f\in\ratfunc$ is of \emph{product type} if
    there are unary functions $U_1(x_1), \dots, U_k(x_k)\colon
    \{0,1\}\to \rationals^{\geq 0}\!$, a polynomial $g\in
    \polyk$ that is a product of binary functions of the form
    $\chi_{=}$ and $\chi_{\neq}$, and a polynomial $s\in \polyk$ such
    that
    \begin{equation}
    \label{eq:product-type}
        f(\xbar) =
            (-1)^{s(\xbar)}
                U_1(x_1) \cdots U_k(x_k)g(\xbar)\,.
    \end{equation}
    The function $f$ is of \emph{product type of degree $d$} if $s$ is of degree
    $d$ and is degree-minimized with respect to $U_1(x_1) \cdots
    U_k(x_k) g(\xbar)$.
\end{definition}

Let $f\in\ratfunc$
be of product type and let $(-1)^{s(\xbar)} U_1(x_1) \cdots U_k(x_k)
g(\xbar)$ be an expression of $f$ as in the definition.  We call a
variable $x_i$ in the representation \emph{determined} if exactly one
of the terms in $g$ is an equality or disequality involving $x_i$,
$U_i(0) = U_i(1) = 1$ and $s$ does not depend on $x_i$.

\begin{example}
\label{ex:prod-type}
    Let $f(x_1,\dots,x_5)$ be the 5-ary function with $f(0,0,1,0,1) =
    8$, $f(0,1,1,0,1) = 10$, $f(1,0,0,1,1) = -12$, $f(1,1,0,1,1) = 15$
    and $f=0$ for all other inputs.  Then $f$ is of product type of
    degree~2, because we can write
    \begin{equation*}
        f(x_1,\dots,x_5) =
            (-1)^{x_1 x_2 \oplus x_1}
            U_1(x_1)\cdots U_5(x_5)
            \chi_{\neq}(x_1, x_3) \chi_{=}(x_1, x_4)\,,
    \end{equation*}
    where $U_1(0) = 2$, $U_1(1) = 3$, $U_2(0) = 4$, $U_2(1) = 5$,
    $U_3(0) = U_3(1) = U_4(0) = U_4(1)=1$, $U_5(0)=0$ and $U_5(1)=1$.
    The variables $x_3$ and $x_4$ are determined.
\end{example}

It is convenient to impose certain restrictions on expressions for
functions of product type.  We say that the expression for $f$ is
\emph{normalized} if the following conditions are met:
\begin{itemize}
\item at least one variable in every equality and disequality term in
    $g$ is determined,
\item if, for some $i$, $U_i(0) = 0$ or $U_i(1) = 0$, then $g$ and
    $s$ do not depend on $x_i$, and
\item $s$ is degree-minimized with respect to $U_1(x_1) \cdots
    U_k(x_k) g(\xbar)$.
\end{itemize}

Note that the expression given in Example~\ref{ex:prod-type} is
normalized: the variables $x_3$ and $x_4$ are determined; $U_5(0) =
0$, so neither $s$ nor $g$ depends on $x_5$; and no sign polynomial of
degree 0 or~1 is equivalent to $x_1 x_2 \oplus x_1$, even with the
flexibility given by the numerous inputs for which $f=0$.

\begin{lemma}
\label{lemma:normalize}
    Every function $f\in\ratfunc$ that is of product type is defined
    by a normalized expression.
\end{lemma}
\begin{proof}
    Let $(-1)^{s(\xbar)} U_1(x_1) \cdots U_k(x_k) g(\xbar)$
    be a non-normalized expression de\-fin\-ing $f$.

    Suppose $g$ contains a term $\chi_{=}(x_i,x_j)$ where neither
    $x_i$ nor $x_j$ is determined.  First, substitute $x_i$ for $x_j$ in
    every other term of $g$ and in $s$.  Replace $U_j$ with the
    function that maps both 0 and~1 to 1 and $U_i$ with the function
    $U_i(x_i)U_j(x_i)$.  The variable $x_j$ is now determined in the resulting
    expression, which still defines $f$.

    Suppose $g$ contains a term $\chi_{\neq}(x_i,x_j)$ where neither
    $x_i$ nor $x_j$ is determined.  We proceed as above but
    substitute $x_i\oplus 1$ for $x_j$.  Having done so, there may
    be terms $\chi_{=}(x_\ell, x_i\oplus 1)$ and $\chi_{\neq}(x_\ell,
    x_i\oplus 1)$; replace these with $\chi_{\neq}(x_\ell, x_i)$ and
    $\chi_{=}(x_\ell, x_i)$, respectively, and similarly for the
    terms with the parameters the other way round.

    Suppose that $U_i(c) = 0$ for some $c\in\{0,1\}$ but $g$ or $s$
    depends on $x_i$.  Since $f$ is zero if $x_i=c$, we may replace
    $x_i$ with $c\oplus 1$ throughout $g$ and $s$.  Performing such a
    replacement in a term of $g$ results in that term becoming a unary
    function, which can be incorporated into the corresponding $U_j$.

    Finally, if $s$ is not degree-minimized, replace it with a
    polynomial in the appropriate variables that is.
\end{proof}

We say that a $k$-ary function $f\colon \{0,1\}^k\to \rationals$ is
\emph{positive pure affine} or of \emph{positive product type} if it
can be written according to Definition~\ref{defn:pure-affine} or
Definition~\ref{defn:product-type}, respectively, but choosing the
sign polynomial $s$ to be identically zero.  Thus, positive pure
affine and positive product type correspond exactly to the definitions
of pure affine and product type used by Dyer et al.\@ for functions
$\{0,1\}^k\to \rationals^{\geq 0}$ \cite{DGJ:weighted-Boolean}.
Observe that, if a function $f\colon\{0,1\}^k\to \rationals^{\geq 0}$
is pure affine (respectively, of product type) then it is positive pure
affine (respectively, of positive product type).  This is because we
must have $s(\xbar)=0$ whenever $f(\xbar)\neq 0$ and, when
$f(\xbar)=0$, we can set $s(\xbar)=0$ without altering the value of
$f$.  Thus, all properties of the functions that Dyer et al.\@ call
``pure affine'' or ``of product type'' in \cite{DGJ:weighted-Boolean}
carry over to non-negative functions that we call pure affine and of
product type, respectively.


\section{The dichotomy}
\label{sec:Dichotomy}

We now give our main result, a complexity dichotomy for Boolean \numCSPclass with
rational weights.  In this section, we prove the tractability of the
polynomial-time cases and comment on our definitions of the classes of
pure affine and product-type functions.  Proving
\FPnumP-completeness of the remaining cases requires considerably
more work and is the subject of the remainder of the paper.

\begin{theorem}
\label{thm:main-theorem}
    Let $\Gamma \subseteq \ratfunc$.  If every function in $\Gamma$ is
    pure affine of degree at most~2, then $\numCSP(\Gamma)$ is in
    \FP.  If every function in $\Gamma$ is of product type of degree
    at most~1, then $\numCSP(\Gamma)$ is in \FP.  Otherwise,
    $\numCSP(\Gamma)$ is \FPnumP-complete.
\end{theorem}
\begin{proof}
    The two polynomial-time cases are covered by Lemmas
    \ref{lemma:easy-affine} and~\ref{lemma:easy-product} in this section.
    If we are not in one of these cases, then $\Gamma$ must contain
    functions $f$ and $g$ (not necessarily distinct) such that $f$ is
    not pure affine of degree at most~2 and $g$ is not of product type
    of degree at most~1.  \FPnumP-completeness follows from
    Lemmas~\ref{lemma:pinning} and~\ref{lemma:hardness-with-pinning}.
\end{proof}

Following from the observations at the end of the previous section, if
we have $\Gamma\subseteq \ratfunc[\geq 0]\!$, then
Theorem~\ref{thm:main-theorem} is equivalent to Theorem~4 of Dyer et
al.~\cite{DGJ:weighted-Boolean}.

It is worth pointing out that we cannot simply dispense with the sign
polynomial in the definitions of pure affine and product type and,
instead, allow the constants and unary functions to take negative
values.  Temporarily call a function $f\colon \{0,1\}^k\to \rationals$
\emph{weakly pure affine} if there is a constant $w\in \rationals$ and
an affine polynomial $g\in \polyk$ such that $f(\xbar)=wg(\xbar)$ and
of \emph{weak product type} if there are unary functions $U_i\colon
\{0,1\}\to \rationals$ and a product $g$ of equalities and
disequalities such that $f(\xbar) = U_1(x_1) \cdots U_k(x_k)
g(\xbar)$.
It is not hard to see that every function that is weakly pure affine
or of weak product type is pure affine or of product type,
respectively.  However, the converse does not hold.  The function
$f(x, y) = (-1)^{xy}$ of Example~\ref{example:xy} above is not weakly pure affine
(there is no rational $w$ such that its range is $\{0, w\}$) and not of weak
product type (it is nowhere zero so there can be no non-trivial
equality or disequality terms and the sign cannot be expressed as a
combination of unary functions).  However, it is trivially pure affine
and of product type (of degree two in both cases).

\begin{lemma}
\label{lemma:easy-affine}
     Let $\Gamma\subseteq\ratfunc$. If every function in $\Gamma$ is pure affine of
     degree at most~2, then $\numCSP(\Gamma)\in \FP$.
\end{lemma}
\begin{proof}
     Let $\Gamma = \{f_1, \dots, f_m\}$, where each $f_i = w_i
     (-1)^{s_i} g_i$ and let $\Gamma' = \{f'_1, \dots, f'_m\}$,
     where each $f'_i = f_i/w_i = (-1)^{s_i} g_i$.  Note that the range
     of each $f'_i$ is included in $\{-1, 0, 1\}$.

     Let $I$ be an instance of $\numCSP(\Gamma)$ and, for each
     $i\in[1,m]$, let $k_i$ be the number of constraints in $I$ that
     involve the function $f_i$.  Let $I'$ be the instance of
     $\numCSP(\Gamma')$ made by replacing each constraint $f(\xbar)$ in
     $I$ with $f'(\xbar)$.  We have
     \begin{equation*}
         Z(I) = Z(I') \!\!\!\prod_{1\leq i\leq m}\!\!\! w_i^{k_i},
     \end{equation*}
     so it suffices to show that we can compute $Z(I')$ in a polynomial
     number of steps.

     If there are $k$ constraints and $n$ variables in $I'\!$, $Z(I')$ is
     a sum of terms of the form
     \begin{equation*}
         \prod_{1\leq j\leq k}\!\! (-1)^{s_{i_j}(\xbar_j)} g_{i_j}(\xbar_j)
             = (-1)^{s(v_1, \dots, v_n)}
                 \!\!\!\prod_{1\leq j\leq k}\!\!\! g_{i_j}(\xbar_j)\,,
     \end{equation*}
     where $s(\vbar) = \sum_{1\leq j\leq k} s_{i_j}(\xbar_j)$.

     We can write $Z(I') = N^+ - N^-\!$, where $N^+$ is the number of
     configurations of the variables of $I$ with weight~1 and $N^-$ is
     the number with weight~$-1$.  Now, $N^+$ is the number of
     solutions of the simultaneous equations
     \begin{equation*}
         g_{i_1}(\xbar_1) = \cdots = g_{i_k}(\xbar_k) = 1
     \end{equation*}
     over \GFtwo that have $s(\vbar)=1$ and $N^-$ is the number of
     solutions with $s(\vbar)=0$.  Since $\Gamma'$ is pure affine of
     degree at most~2, each $g_i$ is linear and $s$ is quadratic.
     Lemma~\ref{lem:quadratic-and-linear} below shows that  the
     number of solutions to such a system of equations can be computed in
     polynomial time.
\end{proof}

\begin{lemma}
\label{lem:quadratic-and-linear}
There is a polynomial-time algorithm for the following problem: given a
multivariate quadratic polynomial $q$ over \GFtwo and $k$
multivariate linear polynomials $\ell_1,\dots, \ell_k$ over
\GFtwo, determine the number of solutions that satisfy $q=0$,
$\ell_1=0,\dots, \ell_k=0$ simultaneously.
\end{lemma}

\begin{proof}
    Suppose $q$ and $\ell_1,\dots, \ell_k$ are in variables $x_1,
    \dots, x_n$ and suppose, without loss of generality, that $\ell_k$
    depends on $x_n$. The polynomial $\ell_k$ evaluates to~0 if, and
    only if, $x_n = h(x_1, \dots, x_{n-1}) = \ell_k \oplus x_n$, where
    $h$ is a linear polynomial in $x_1,\dots,x_{n-1}$. Substitute $h$
    for $x_n$ in $q$ and $\ell_1, \dots, \ell_{k-1}$ to obtain $q'$
    and $\ell'_1, \dots, \ell'_{k-1}$, respectively. The number of
    solutions that satisfy $q=0$, $\ell_1=0,\dots, \ell_k=0$ is the
    same as the number of solutions that satisfy $q'=0$,
    $\ell'_1=0,\dots, \ell'_{k-1}=0$, which may be found
    recursively. We process recursively until the system of equations
    contains one
    quadratic equation and no linear equations, or only linear
    equations.  The number of solutions to a quadratic polynomial
    equation over \GFtwo can be computed in polynomial
    time~\cite{EK1990:equations,LN1997:fin-fields}. The number of
    solutions of a system of linear equations over \GFtwo can be
    computed by Gaussian elimination in polynomial time.
\end{proof}

The case where every function in $\Gamma$ is of product type of degree
at most~1 is essentially the same as the corresponding case for
non-negative functions \cite{DGJ:weighted-Boolean} but we give a full
proof for completeness.

\begin{lemma}
\label{lemma:easy-product}
    Let $\Gamma\subseteq\ratfunc$. If every function in $\Gamma$ is of product type
    of degree at most~1, then $\numCSP(\Gamma)\in \FP$.
\end{lemma}
\begin{proof}
    Observe that, since each function $f\in \Gamma$ is of product type
    of degree at most~1, each can be written in the form
    \begin{align*}
        f(\xbar)
            &= (-1)^{x_{i_1} + \dots + x_{i_\ell} + c}
                   U_1(x_1) \cdots U_k(x_k) g(\xbar)                  \\
            &= (-1)^{x_{i_1}} \cdots (-1)^{x_{i_\ell}} (-1)^c
                   U_1(x_1) \cdots U_k(x_k) g(\xbar)\,,
    \end{align*}
    for some $c\in\{0, 1\}$.  Thus, we can, instead, write $f(\xbar) =
    U'_1(x_1) \cdots U'_k(x_k) \cdot h(\xbar)$ where each $U'_i$ is a
    function $\{0,1\}\to \rationals$ instead of $\{0,1\}\to
    \rationals^{\geq 0}\!$.  The remainder of the proof is the same as
    the corresponding case for non-negative functions.

    Let $I$ be an instance of $\numCSP(\Gamma)$, with variables $V$.
    Let $\approx$ be the finest equivalence relation over $V$ such
    that $v_i\approx v_j$ if $i = j$ or some constraint in $I$
    requires that either $v_i = v_j$ or $v_i\neq v_j$.  We process
    each equivalence class in turn, independently of the others.

    Let $S\subseteq V$ be an equivalence class of $\approx$.  If there
    is no assignment to the variables in $S$ that satisfies the
    equalities and disequalities in $I$'s constraints, then $Z(I)=0$
    and we are done.  Otherwise, $S$ must have a partition into sets
    $S_0$ and $S_1$ so that each variable in $S_0$ must have the same
    value and each variable in $S_1$ (which may be empty) must have
    the opposite value.  The variables in $S$ contribute one weight,
    say $\alpha$, to $Z(I)$ if the variables in $S_0$ are set to~0 and
    another weight, say $\beta$, if they are set to~1.  Thus, we can
    write $Z(I) = (\alpha + \beta)Z'(I)$, where $Z'(I)$ is the
    partition function $Z(I)$ with all terms involving the variables
    in $S$ deleted.  We may then proceed to factor out the next
    equivalence class.
\end{proof}


\section{Useful reductions}
\label{sec:Reductions}

In this section, we give several reductions that are useful for
proving hardness of weighted Boolean \numCSP{}s.


\subsection{Pinning}

Let $\delta_0$ and $\delta_1$ be the unary functions defined as
\begin{align*}
    \delta_0(0) &= 1 & \delta_1(0) = 0 \\
    \delta_0(1) &= 0 & \delta_1(1) = 1
\end{align*}
These functions are referred to as \emph{pinning} functions, since a
constraint $\delta_c(x)$ ``forces'' the variable $x$ to take value $c$
by giving weight zero to any configuration with $x\neq c$.  The proof
of the following lemma is identical to the proof of
\cite[Lemma~8]{DGJ:weighted-Boolean}, except that the condition
``$f(x) > f(\xbar) \geq 0$'' in the first sentence of Case~2 needs to
be replaced with ``$f(x) \neq f(\xbar)$''.

\begin{lemma}
\label{lemma:pinning}
    For every $\Gamma\subseteq \ratfunc$, $\numCSP(\Gamma \cup
    \{\delta_0, \delta_1\})\reducesto \numCSP(\Gamma)$.
\end{lemma}


\subsection{Arity reduction}

Given a $k$-ary function $f\in\ratfunc$ and $i\in[1,k]$, the function
obtained by \emph{projecting out the $i$th variable} is
\begin{equation*}
    g(x_1, \dots, x_{k-1}) = \!\!\!\sum_{y\in\{0,1\}}\!\!\!
                       f(x_1, \dots, x_{i-1}, y, x_i, \dots, x_{k-1})\,.
\end{equation*}

The following is a special case of
\cite[Lemma~6]{DGJ:weighted-Boolean}.  Although that Lemma is stated
only for classes of non-negative rational functions, the proof does
not rely on this.

\begin{lemma}
\label{lemma:projection}
    Let $\Gamma\subseteq\ratfunc$, let $f\in \Gamma$ and let $g$ be
    defined by projecting out a variable of $f$.  $\numCSP(\Gamma \cup
    \{g\}) \reducesto \numCSP(\Gamma)$.
\end{lemma}

The \emph{contraction} of a ternary function $f\in\ratfunc$
is the function
\begin{equation*}
    g(x_1, x_2) = \negquad\sum_{y,z\in\{0,1\}}\negquad
                                            f(x_1, y, z) f(y, z, x_2)\,.
\end{equation*}
(In principle, we could define contractions in terms of any sequence
of function arguments but we only use the version defined
here.)

\begin{lemma}
\label{lemma:contract}
    Let $\Gamma\subseteq\ratfunc$ and let $g$ be the contraction of
    some $f\in \Gamma$.  $\numCSP(\Gamma \cup \{g\}) \reducesto
    \numCSP(\Gamma)$.
\end{lemma}
\begin{proof}
    Replace each constraint $C$ of the form $g(x,y)$ with the two
    constraints $f(x, x_C, y_C)$ and $f(x_C, y_C, y)$, where $x_C$ and
    $y_C$ are new variables, used only in these two constraints.
\end{proof}


\subsection{Arithmetic techniques}

For a constant $q\in\rationals$ and a function $f\in\ratfunc$, write
$qf$ for the function that maps $\xbar$ to $qf(\xbar)$.

\begin{lemma}
\label{lemma:factor}
    Let $f\in\Gamma\subseteq \ratfunc$ and let $q\neq 0$ be rational.
    $\numCSP(\Gamma \cup \{qf\}) \reducesto \numCSP(\Gamma)$.
\end{lemma}
\begin{proof}
    Let $I$ be an instance of $\numCSP(\Gamma \cup \{qf\})$ and let
    $I'$ be the instance of $\numCSP(\Gamma)$ made by replacing every
    constraint $qf(\xbar)$ in $I$ with $f(\xbar)$.  $Z(I) = q^m
    Z(I')$, where $m$ is the number of $qf$-constraints in $I$.
\end{proof}

Given a constraint language $\Gamma\!$, let $\Gamma^2$ be the constraint
language that replaces every function $f(\xbar)$ with the function
$(f(\xbar))^2\!$.  The following lemma is immediate from the
observation that an instance of $\numCSP(\Gamma^2)$ can be converted to
one of $\Gamma$ with the same partition function just by including an
extra copy of each constraint.

\begin{lemma}
\label{lemma:squaring}
    $\numCSP(\Gamma^2)\reducesto \numCSP(\Gamma)$.
\end{lemma}

Further, if $\Gamma\subseteq \ratfunc$, then $\Gamma^2\subseteq
\ratfunc[\geq 0]\!$.  This fact and the following lemma allow us to
re-use results on those functions from \cite{DGJ:weighted-Boolean}.

\begin{lemma}
\label{lemma:square-preserve}
    $f \in \ratfunc$ is pure affine (respectively, of product type)
    if, and only if, $f^2 \in \ratfunc[\geq 0]$ is pure affine
    (respectively, of product type).
\end{lemma}
\begin{proof}
    Let $f \in \ratfunc$ be $k$-ary.  It is clear that, if $f$ is pure
    affine (respectively, of product type), then so is $f^2$; we show
    the converse.  We assume that $f^2$ is not identically zero as
    this case is trivial.

    First, suppose $f^2(\xbar)$ is pure affine and equal to $w^2
    g(\xbar)$ as in Definition~\ref{defn:pure-affine}.  There is a
    polynomial $s(\xbar)$ that assigns the correct sign to each input
    such that $f(\xbar) = w(-1)^{s(\xbar)} g(\xbar)$.  Therefore $f$
    is pure affine.

    Now, suppose $f^2(\xbar) = U_1(x_1) \cdots U_k(x_k) g(\xbar)$,
    as in Definition~\ref{defn:product-type}, is of product type.  For
    each $i\in[1,k]$, let $U'_i(x_i) = \sqrt{U_i(x_i)}$.  The
    functions $U'_i$ are not necessarily rational but we certainly
    have
    \begin{equation}
    \label{eq:irrational-unary-functions}
        f(\xbar)=(-1)^{s(\xbar)} U'_1(x_1) \cdots U'_k(x_k) g(\xbar)\,,
    \end{equation}
    for some suitable polynomial $s$, as before.  By the arguments
    of Lemma~\ref{lemma:normalize}, which do not depend on the
    rationality of the functions $U'_i$, we may assume that this is a
    normalized expression for $f$, except for the possible
    irrationality of the $U'_i$.

    We now describe how the functions $U'_i$ can be replaced by
    rational functions, keeping the expression for $f$ normalized. The
    function $f$ is not identically zero so there is a tuple $\abar
    \in \{0,1\}^k$ such that $f(\abar) \neq 0$. Since $f \in
    \ratfunc$, $f(\abar)$ is rational.  For $i \in [1,k]$, let
    $U''_i(x_i) = U'_i(x_i) / U'_i(a_i)$. Then
    \begin{equation}
    \label{eq:rational-unary-functions}
        f(\xbar) = |f(\abar)|(-1)^{s(\xbar)}
U''_1(x_1) \cdots U''_k(x_k) g(\xbar)\,.
    \end{equation}
    The expression for $f$ in
    (\ref{eq:rational-unary-functions}) is not necessarily
    normalized because of the factor $|f(\abar)|$; however, as we will
    see next, the functions $U''_i$ are rational. Once we have
    established this fact we will see that the factor $|f(\abar)|$
    (which is rational) can be included in one of the unary functions
    $U''_i$, giving us a normalized expression for $f$.

    Note that, for $i \in [1,k]$, $U''_i(a_i) = 1$, which is
    rational. Therefore we need to show that $U''_i(a_i \oplus 1)$ is
    rational. Observe that, for each $i \in [1,k]$ for which $x_i$ is
    determined in (\ref{eq:irrational-unary-functions}), we have
    $U''_i(0) = U''_i(1) = 1$. We now show that, for each $i \in
    [1,k]$ for which $x_i$ is not determined, $U''_i(a_i \oplus 1)$ is
    rational. Suppose $U''_i(a_i \oplus 1) \neq 0$. Let $\abar_i \in
    \{0,1\}^k$ be the tuple obtained from $\abar$ by replacing $a_i$
    with $a_i \oplus 1$ and replacing $a_j$ with $a_j \oplus 1$ for
    every determined variable $x_j$ that occurs together with $x_i$ in
    an equality or disequality function of $g$. Thus, $g(\abar_i) = 1$
    and $|f(\abar_i)| = |f(\abar)| U''_i(a_i \oplus 1) > 0$. Since $f
    \in \ratfunc$ and $f(\abar)$ is rational, $U''_i(a_i \oplus 1)$ is
    rational.

    Finally we notice that the factor $|f(\abar)|$ in
    (\ref{eq:rational-unary-functions}) can be absorbed by
    any of the unary functions $U''_i$ for which $x_i$ is not a
    determined variable in the expression for $f$ in
    (\ref{eq:irrational-unary-functions}). If all variables
    are determined then we note that $|f(\abar)| = 1$ and we can
    disregard it completely. We finally conclude that $f(\xbar)$ is of
    product type.
\end{proof}


\subsection{Matrix techniques}

Given a $k\times k$ rational matrix, $A=(A_{ij})$, and a directed
multigraph $G=(V,E)$, which may have loops, let
\begin{equation*}
    Z_A(G) = \negquad\sum_{\sigma\colon V \to [1,k]}
                 \prod_{(x,y) \in E}\!\!\! A_{\sigma(x) \sigma(y)}\,.
\end{equation*}
The problem of computing $Z_A(G)$ for a given input graph $G$ is
denoted by $\Eval{A}$.  Bulatov and Grohe have given the complexity of
$\Eval{A}$ for any symmetric matrix $A$ with non-negative entries
\cite{BG2005:partition}.  Here we only need the following special
case.

\begin{lemma}
\label{lemma:Bulatov-Grohe}
    Let $A$ be a symmetric $2\times 2$ matrix with non-negative
    rational entries.  If $A$ has rank~2 and at most one entry of $A$
    is zero then $\Eval{A}$ is \numP-hard.
\end{lemma}

For any $k\times k$ rational matrix $A$, $\Eval{A}$ is just the same
thing as $\numCSP(f)$ for an appropriate binary function $f$ over a
domain of size $k$.  In particular, then, $2\times 2$ matrices
correspond to binary Boolean functions.

\begin{lemma}
\label{lemma:Eval-to-numCSP}
    Let $f \in \ratfunc$ be a binary function and let $A$ be the
    matrix
    \begin{equation*}
        A = \begin{pmatrix}
                f(0,0) & f(0,1)\\
                f(1,0) & f(1,1)
            \end{pmatrix}.
    \end{equation*}
    Then $\Eval{A} \equivto \numCSP(f)$.
\end{lemma}
\begin{proof}
    Given an instance graph $G=(V,E)$ of $\Eval{A}$, let $I$ be the
    instance of $\numCSP(f)$ with variables $V$ that has a
    constraint $f(x,y)$ for every edge $(x,y)$ in $E$. Thus $Z_A(G) =
    Z(I)$.
\end{proof}

While Lemma~\ref{lemma:Eval-to-numCSP} applies to all rational
functions $f$, Lemma~\ref{lemma:Bulatov-Grohe} can only be used if the
resulting matrix is both symmetric and non-negative.  The following
lemma, essentially due to Dyer and Greenhill \cite{DG2000:graph-homo}
will allow us to transform the matrix corresponding to a function $f$
into a symmetric, non-negative matrix.  For a matrix $A = (A_{ij})$,
we write $A^{(2)}$ for the matrix $(A_{ij}^2)$.

\begin{lemma}
\label{lemma:matrix-op}
    For any rational square matrix $A$, the problems $\Eval{A^{(2)}}$,
    $\Eval{A \transp{A}}$, $\Eval{\transp{A}A}$ and $\Eval{A^2}$ are
    polynomial-time Turing-redu\-cible to $\Eval{A}$.
\end{lemma}
\begin{proof}
    For any graph $G$,
    \begin{itemize}
    \item $Z_{A^{(2)}}(G) = Z_A(G_1)$, where $G_1$ is the multigraph
        formed by replacing each edge of $G$ with two parallel edges;
    \item $Z_{A\transp{A}}(G) = Z_A(G_2)$, where $G_2$ is the graph
        obtained by introducing a new vertex $v_e$ for each edge
        $e=(x,y)\in G$ and replacing $e$ with the edges $(x,v_e)$ and
        $(y,v_e)$;
    \item $Z_{\transp{A}A}(G) = Z_A(G_3)$, where $G_3$ is made in the
        same way as $G_2$ but replacing $e$ with $(v_e, x)$ and $(v_e,
        y)$;
    \item $Z_{A^2}(G) = Z_A(G_4)$, where $G_4$ is made in the same way
        as $G_2$ but replacing $e$ with $(x, v_e)$ and $(v_e, y)$.
    \qedhere
    \end{itemize}
\end{proof}


\section{High-degree pure affine functions}
\label{sec:Hard-affine}

We have seen that there is a polynomial-time algorithm for
$\numCSP(\Gamma)$ if every function in $\Gamma$ is pure affine of
degree at most two.  We now show that computing partition functions of
pure affine functions of higher degree is \numP-hard.  The main
result of this section is the following lemma.

\begin{lemma}
\label{lemma:pure-affine-high-degree}
    If $f \in \ratfunc$ is pure affine of degree at least three, then
    $\numCSP(f)$ is \numP-hard.
\end{lemma}

We first consider the restricted case $f(x,y,z)=(-1)^{s(x,y,z)}\!$,
for ternary functions $s$ of degree exactly~3 and then show that the
case $f(\xbar) = (-1)^{s(\xbar)}$ of degree-3 functions of arbitrary
arity greater than three follows.  Finally, we prove
Lemma~\ref{lemma:pure-affine-high-degree}.

\begin{table}[t]
\centering\renewcommand{\arraystretch}{1.15}
\begin{tabular}{|l|l|c|c|}
    \hline
    $s(x,y,z)$           & Method          &           $A$       &             $A'$   \rule{0pt}{16pt}     \\
    \hline
    $xyz$                & Project out $z$ & \smat{ 2& 2\\ 2& 0} & \smat{ 64& 16\\ 16& 16} \\
    $xyz+x$              & Project out $z$ & \smat{ 2& 2\\-2& 0} & \smat{ 64& 16\\ 16& 16} \\
    $xyz+x+y$            & Project out $z$ & \smat{ 2&-2\\-2& 0} & \smat{ 64& 16\\ 16& 16} \\
    $xyz+x+y+z$          & Contract        & \smat{ 4&-2\\-2& 4} & \smat{ 400& 256\\ 256& 400} \\
    $xyz+xy$             & Project out $z$ & \smat{ 2& 2\\ 2& 0} & \smat{ 64& 16\\ 16& 16} \\
    $xyz+xy+x$           & Project out $z$ & \smat{ 2& 2\\-2& 0} & \smat{ 64& 16\\ 16& 16} \\
    $xyz+xy+x+y$         & Project out $z$ & \smat{ 2&-2\\-2& 0} & \smat{ 64& 16\\ 16& 16} \\
    $xyz+xy+z$           & Project out $y$ & \smat{ 2&-2\\ 0&-2} & \smat{ 64& 16\\ 16& 16} \\
    $xyz+xy+x+z$         & Project out $y$ & \smat{ 2&-2\\ 0& 2} & \smat{ 64& 16\\ 16& 16} \\
    $xyz+xy+x+y+z$       & Contract        & \smat{ 2&-4\\ 0& 2} & \smat{400& 64\\ 64& 16} \\
    $xyz+xy+xz$          & Project out $z$ & \smat{ 2& 2\\ 0&-2} & \smat{ 64& 16\\ 16& 16} \\
    $xyz+xy+xz+x$        & Project out $z$ & \smat{ 2& 2\\ 0& 2} & \smat{ 64& 16\\ 16& 16} \\
    $xyz+xy+xz+y$        & Project out $z$ & \smat{ 2&-2\\ 0& 2} & \smat{ 64& 16\\ 16& 16} \\
    $xyz+xy+xz+x+y$      & Project out $z$ & \smat{ 2&-2\\ 0&-2} & \smat{ 64& 16\\ 16& 16} \\
    $xyz+xy+xz+y+z$      & Contract        & \smat{ 2&-4\\ 0& 2} & \smat{400& 64\\ 64& 16} \\
    $xyz+xy+xz+x+y+z$    & Project out $x$ & \smat{ 0&-2\\-2& 2} & \smat{ 16& 16\\ 16& 64} \\
    $xyz+xy+xz+yz$       & Contract        & \smat{ 4&-2\\-2& 4} & \smat{400&256\\256&400} \\
    $xyz+xy+xz+yz+x$     & Project out $x$ & \smat{ 0& 2\\ 2&-2} & \smat{ 16& 16\\ 16& 64} \\
    $xyz+xy+xz+yz+x+y$   & Project out $y$ & \smat{ 0& 2\\-2& 2} & \smat{ 16& 16\\ 16& 64} \\
    $xyz+xy+xz+yz+x+y+z$ & Project out $z$ & \smat{ 0&-2\\-2&-2} & \smat{ 16& 16\\ 16& 64} \\
    \hline
\end{tabular}
\caption{The twenty ternary degree-3 polynomials considered in
    Lemma~\ref{lemma:degree-3-hard}, with the methods used to prove them
    hard and the corresponding matrices.}
\label{table:deg3-cases}
\end{table}

\begin{lemma}
\label{lemma:degree-3-hard}
    Let $f(x, y, z) = (-1)^{s(x, y, z)}$ where $s\in\polyk[3]$ is of
    degree~3.  $\numCSP(f)$ is \numP-hard.
\end{lemma}
\begin{proof}
    Since $s$ is of degree~3, it must contain the term $xyz$.  Note
    that $\numCSP((-1)^{s(x, y, z)})$ is equivalent to
    $\numCSP((-1)^{s(x, y,z)+ 1})$ under Turing reductions,
    since $Z(I) = (-1)^m Z(I')$  where $I$ and
    $I'$ are instances of the two problems with the same $m$
    constraints.  Therefore, we may assume that $s$ does not contain
    the constant term~1.

    Given this assumption, the terms of $s$ are $xyz$ and some subset
    of the terms $xy$, $yz$, $zx$, $x$, $y$ and~$z$.  By symmetry
    between the variables, there are twenty cases to consider, listed
    in Table~\ref{table:deg3-cases}.  Each case is proven \numP-hard
    by either projecting out a variable or contracting, as detailed in
    the table.

    For each polynomial $s$ listed in the table, let $f(x,y,z) = (-1)^{s(x,y,z)}\!$.
    Note that $f$ has the same value when $s$ is evaluated over $\ints$
    as it does when $s$ is evaluated over \GFtwo, so we need not distinguish
    between $+$ and $\oplus$. The operation given (projecting out a
    variable or contracting) produces a new function $f'$ in two
    variables which we will call $x$ and $y$.  By
    Lemma~\ref{lemma:projection} (projection), or
    Lemma~\ref{lemma:contract} (contraction), $\numCSP(f')\reducesto
    \numCSP(f)$.  Further, by Lemma~\ref{lemma:Eval-to-numCSP},
    $\Eval{A} \reducesto \numCSP(f')$, where
    \begin{equation*}
        A = \begin{pmatrix}
                f'(0,0) & f'(0,1) \\
                f'(1,0) & f'(1,1)
            \end{pmatrix}
    \end{equation*}
    is given in the table.  Let $A' = (A\transp{A})^{(2)}\!$.  For any
    rational matrix $A$, the corresponding $A'$ is symmetric and
    non-negative and, by Lemma~\ref{lemma:matrix-op}, $\Eval{A'}
    \reducesto \Eval{A}$.  All of the matrices $A'$ given in the table
    have rank~2 and no zero entries so, by
    Lemma~\ref{lemma:Bulatov-Grohe}, $\Eval{A'}$ is \numP-hard.
\end{proof}

\begin{lemma}
\label{lemma:big-degree-hard}
    Let $\xbar = x_1\dots x_k$ for some $k>3$ and let $f(\xbar) =
    (-1)^{s(\xbar)}$ for some $s\in \polyk$ of degree at least~3.
    $\numCSP(f)$ is \numP-hard.
\end{lemma}
\begin{proof}
    Renaming variables if necessary, we may assume that one of the
    terms of least degree greater than or equal to three in $s$ is $x_1\cdots
    x_\ell$ for some $\ell$ with $3\leq \ell\leq k$.  Let $c_4 =
    \cdots = c_\ell = 1$ and $c_{\ell+1} = \cdots = c_k = 0$ and let
    $s'(x_1, x_2, x_3) = s(x_1, x_2, x_3, c_4, \dots, c_k)$ and
    $f'(x_1, x_2, x_3) = (-1)^{s'(x_1, x_2, x_3)}\!$.

    The degree of $s'$ is~3 since it has only three variables and
    includes exactly one term $x_1 x_2 x_3 1\cdots 1 = x_1 x_2 x_3$.
    Therefore, $\numCSP(f')$ is \numP-hard by the previous lemma.

    It remains to show that $\numCSP(f')\reducesto \numCSP(f)$.  To
    see this, let $I'$ be any instance of $\numCSP(f')$.  We create an
    instance $I''$ of $\numCSP(\{f, \delta_0, \delta_1\})$ such that
    $Z(I'') = Z(I')$ as follows, and the result is then immediate from
    Lemma~\ref{lemma:pinning}.  Let $z_4, \dots, z_k$ be new
    variables.  Let $I''$ have the constraints $\delta_1(z_4), \dots,
    \delta_1(z_\ell)$, $\delta_0(z_{\ell+1}), \dots, \delta_0(z_k)$
    and, for each constraint $f'(y_1, y_2, y_3)$ in $I'\!$, the
    constraint $f(y_1, y_2, y_3, z_4, \dots, z_k)$.
\end{proof}

We now prove the main result of this section, namely that $\numCSP(f)$
is \numP-hard if $f$ is pure affine of degree at least three.

\begin{proof}[Proof of Lemma~\ref{lemma:pure-affine-high-degree}]
    Let $f(x_1, \dots, x_k)$ be pure affine of degree at least three.
    Thus, we may write $f(\xbar) = w(-1)^{s(\xbar)}g(\xbar)$, where
    $w>0$, $g\in \polyk$ is affine and $s\in \polyk$ is
    degree-minimized with respect to $g$ and has degree at least three.
    By Lemma~\ref{lemma:factor}, we may assume that $w=1$.

    Since $g$ is affine, we may write
    \begin{equation*}
        f(\xbar) = (-1)^{s(\xbar)} \!\!\prod_{i \in[1,m]}\!\! g_i(\xbar),
    \end{equation*}
    where each $g_i\in\polyk$ is linear.  We show that $f$ is
    \numP-hard by induction on $m$.  The base case, $m=0$, is
    Lemma~\ref{lemma:big-degree-hard}.

    For the inductive step $m > 0$, we may assume without loss of
    generality that $g_m$ depends on $x_k$.  If $f(\xbar)\neq 0$, we
    must have $g_m(\xbar)=1$ and, therefore, $x_k = g_m(\xbar) \oplus x_k \oplus
    1$.  Note that $g_m(\xbar) \oplus x_k \oplus 1$ does not depend on $x_k$.
    Let $g'_1, \dots, g'_{m-1}, s'\in \polyk[k-1]$ be the polynomials
    that result from substituting $g_m(\xbar) \oplus x_k \oplus 1$ for $x_k$ in
    $g_1, \dots, g_{m-1}$ and $s$, respectively.

    Since $s'(\xbar) = s(\xbar)$ whenever $f(\xbar)\neq 0$, we have
    $f(\xbar) = (-1)^{s'(\xbar)} g(\xbar)$.  Because $s$ is
    degree-minimized with respect to $g$, $s'$ must have the same
    degree as $s$.

    Let
    \begin{equation*}
        f'(x_1, \dots, x_{k-1}) =
            (-1)^{s'(\xbar)} \negquad\!\prod_{i\in[1,m-1]}\negquad\! g'_i(\xbar)\,.
    \end{equation*}
    Suppose that $s'$ is not degree-minimized with respect to
    $g'(\xbar) = \prod_i g'_i(\xbar)$.  Then there is another
    polynomial $s''$ of strictly lower degree such that $f'(\xbar) =
    (-1)^{s''(\xbar)} g'(\xbar)$.  But then, we have $f(\xbar) =
    (-1)^{s''(\xbar)} g_m(\xbar)g'(\xbar) = (-1)^{s''(\xbar)}
    g(\xbar)$, contradicting degree-minimality of $s$.  Therefore,
    $s'$ is degree-minimized with respect to $g'\!$.  Further, $s$ and
    $s'$ have the same degree, so $\numCSP(f')$ is \numP-hard by the
    inductive hypothesis.

    It remains to show that $\numCSP(f')\reducesto \numCSP(f)$.  Let
    $I'$ be an instance of $\numCSP(f')$ and let $I$ be the instance
    of $\numCSP(f)$ that has a constraint $f(x_1, \dots, x_{k-1},
    x_C)$ for every constraint $C= f'(x_1, \dots, x_{k-1})$ in $I'\!$.
    Then $Z(I) = Z(I')$ and we are done.
\end{proof}


\section{High-degree product-type functions}
\label{sec:Hard-product}

We now construct the machinery for the remaining hard case: functions
of product type of degree two or more that are not pure affine of
degree two.

For any $\lambda\in\rationals$, we write $\Theta_\lambda(x)$ for the
function $\Theta_\lambda(0) = 1$, $\Theta_\lambda(1) = \lambda$.  (In
\cite{DGJ:weighted-Boolean}, these functions are written $U_\lambda$
but we wish to avoid the potential for confusion with the functions
$U_1, \dots, U_k$ used to define a $k$-ary function of product type.)

The main result of this section is the following lemma.

\begin{lemma}
\label{lemma:product-type-high-degree}
    Let $f\in \ratfunc$ be of product type of degree at least
    two. Then, $\numCSP(\{f, \Theta_\lambda\})$ is \numP-hard for
    any positive rational $\lambda\neq 1$.
\end{lemma}

If $f$ is both of product type of degree two and pure affine of degree
two, then $\numCSP(f)$ is computable in polynomial time by
Lemma~\ref{lemma:easy-affine}.  In the following section, we will show
that, for all other functions of product type of degree two or more,
we have $\numCSP(\{f, \Theta_\lambda\}) \equivto \numCSP(f)$ so
Lemma~\ref{lemma:product-type-high-degree} is sufficient for our
needs, even though it appears, at first sight, to be weaker than the
desired result.

As in the previous section, we first consider simplified cases.

\begin{lemma}
\label{lemma:prod-type-two-variables}
    Let $f$ be of product type of degree at least two.  There are
    non-zero rationals $\alpha$ and $\beta$ such that $\numCSP(f')
    \reducesto \numCSP(f)$, where $f'(x,y) = (-1)^{xy}
    \Theta_{\alpha}(x) \Theta_{\beta}(y)$.
\end{lemma}
\begin{proof}
    Let $(-1)^{s(\xbar)} U_1(x_1) \cdots U_k(x_k) g(\xbar)$ be a
    normalized expression defining $f$.

    We may assume, renaming variables if necessary, that
    \begin{equation*}
        s(\xbar) = x_1x_2 p(x_3, \dots, x_k) + q(x_1, \dots, x_k)\,,
    \end{equation*}
    where $p$ and $q$ are polynomials in the stated variables, $p$ is
    not identically zero and $q$ contains no term that has $x_1x_2$ as
    a factor.  Let $X$ be the set of variables on which $s$ depends.
    Because $x_1\in X$, there must be an assignment $\sigma\colon
    X\to\{0,1\}$ such that $s(0, \sigma(x_2), \dots, \sigma(x_k)) \neq
    s(1, \sigma(x_2), \dots, \sigma(x_k))$.  We may assume that
    $s(\sigma(\xbar)) = 1$.

    Now let $Y$ be the set of variables on which $g$ depends.  Since
    the expression is normalized, at least one variable in each term
    $\chi_{=}(x_i, x_j)$ or $\chi_{\neq}(x_i, x_j)$ is determined and
    no determined variable appears in $s$.  Therefore, we can extend
    $\sigma$ to an assignment $\sigma'\colon X\cup Y\to \{0, 1\}$ such
    that $s(\sigma'(\xbar)) = g(\sigma'(\xbar)) = 1$.

    Further, for every $i$ with $x_i\in X\cup Y$, $U_i(0)$ and
    $U_i(1)$ are both non-zero.  For each $x_i\notin (X\cup Y)$, we
    must have $U_i(0)\neq 0$ or $U_i(1)\neq 0$ or both; otherwise, $f$
    is identically zero (and, thus, of product type of degree zero).
    Therefore, we can extend $\sigma'$ to an assignment
    $\sigma''\colon \{x_1, \dots, x_k\}\to \{0, 1\}$ such that
    $f(\sigma''(\xbar))\neq 0$.

    Finally, suppose that $f'(x,y) = f(x, y, \sigma''(x_3), \dots,
    \sigma''(x_k))$.  Then clearly $\numCSP(f')\reducesto \numCSP(\{f,
    \delta_0, \delta_1\})$ so, by Lemma~\ref{lemma:pinning},
    $\numCSP(f')\reducesto \numCSP(f)$.

    There are constants $w\in\rationals$ and $a, b, c\in \{0, 1\}$
    such that
    \begin{align*}
        f'(x, y)
            &= w (-1)^{xy + ax + by + c} U_1(x) U_2(y)                \\
            &= w (-1)^c (-1)^{xy} U'_1(x) U'_2(y)\,,
    \end{align*}
    where $U'_1(x) = (-1)^{ax} U_1(x)$ and $U'_2(y) = (-1)^{by}
    U_2(y)$.  We can now put $\alpha = U'_1(1)/U'_1(0)$ and $\beta =
    U'_2(1)/U'_2(0)$, giving
    \begin{equation*}
        f'(x, y) = w U'_1(0) U'_2(0) (-1)^c (-1)^{xy}
                   \Theta_{\alpha}(x) \Theta_{\beta}(y)\,.
    \end{equation*}
    By Lemma~\ref{lemma:factor}, we can discard the constant
    factor $w U'_1(0) U'_2(0) (-1)^c\!$.
\end{proof}

\begin{lemma}
\label{lemma:prod-type-hard-two-variables}
    If $f(x,y) = (-1)^{x y} \Theta_\alpha(x)
    \Theta_\beta(y)$, where $\alpha \in \rationals\setminus\{-1,
    0,1\}$, $\beta \in \rationals \setminus\{0\}$, then $\numCSP(f)$
    is \numP-hard.
\end{lemma}
\begin{proof}
    Let
    \begin{equation*}
        A = \begin{pmatrix}
                f(0,0) & f(0,1) \\
                f(1,0) & f(1,1)
            \end{pmatrix}
          = \begin{pmatrix}
                1      & \beta        \\
                \alpha & -\alpha\beta
            \end{pmatrix}
    \end{equation*}
    and let
    \begin{equation*}
        B = (\transp{A}A)^{(2)} =
            \begin{pmatrix}
                1 + \alpha^2         & \beta (1 - \alpha^2)   \\
                \beta (1 - \alpha^2) & \beta^2 (1 + \alpha^2)
            \end{pmatrix}^{(2)}.
    \end{equation*}
    Since $\beta\neq 0$ and $\alpha^2\neq 1$, every entry of $B$ is
    positive.  We have
     \begin{equation*}
        |B| = (1+\alpha^2)^4 \beta^4 - (1-\alpha^2)^4 \beta^4
            = 8\alpha^2 \beta^4 (1+\alpha^4)
            > 0\,.
    \end{equation*}

    Therefore, $B$ has rank two and hence $\Eval{B}$ is \numP-hard by
    Lemma~\ref{lemma:Bulatov-Grohe}.  By Lemmas
    \ref{lemma:Eval-to-numCSP} and~\ref{lemma:matrix-op},
    $\Eval{B}\reducesto \Eval{A} \equivto \numCSP(f)$.
\end{proof}

We now prove Lemma~\ref{lemma:product-type-high-degree}, namely that,
if $f$ is of product type of degree at least two, then $\numCSP(\{f,
\Theta_\lambda\})$ is \numP-hard for any positive rational
$\lambda\neq 1$.

\begin{proof}[Proof of Lemma~\ref{lemma:product-type-high-degree}]
    Let $f\in\ratfunc$ be of product type of degree at least~2 and let
    $\lambda\in \rationals^{>0} \setminus \{1\}$.  By
    Lemma~\ref{lemma:prod-type-two-variables} there are non-zero
    rational constants $\alpha$ and $\beta$ such that $\numCSP(g)
    \reducesto \numCSP(f)$, where
    \begin{equation*}
        g(x, y) = (-1)^{x y} \Theta_\alpha(x) \Theta_\beta(y)\,.
    \end{equation*}

    If at most one of $\alpha$ and $\beta$ is 1 or $-1$, then
    $\numCSP(g)$ is \numP-hard by
    Lemma~\ref{lemma:prod-type-hard-two-variables} and we are done.
    Otherwise, we have $\alpha, \beta \in \{-1, 1\}$.  Let
    \begin{equation*}
        g'(x, y) = (-1)^{x y}
                       \Theta_{\alpha\lambda}(x) \Theta_\beta(y)\,.
    \end{equation*}
    $\numCSP(g')$ is \numP-hard by
    Lemma~\ref{lemma:prod-type-hard-two-variables}, since
    $\alpha\lambda\notin \{-1, 0, 1\}$.  It just remains to show that
    $\numCSP(g')\reducesto \numCSP(\{g, \Theta_\lambda\})$ but this
    is easy: given an instance of $\numCSP(g')$, replace every
    constraint $g'(x,y)$ by the pair of constraints $g(x,y)$ and
    $\Theta_\lambda(x)$.
\end{proof}


\section{Proving the dichotomy}
\label{sec:Proof}

We now have all the tools we need to prove the remaining side of the
dichotomy, namely that, unless either every $f\in\Gamma$ is pure
affine of degree at most two or every $f$ is of product type of degree
at most one, then $\numCSP(\Gamma)$ is \numP-hard.

\begin{lemma}
\label{lemma:not-affine}
    If $f\in\ratfunc$ does not have affine support, then $\numCSP(f)$
    is \numP-hard.
\end{lemma}
\begin{proof}
    $f^2\in\ratfunc[\geq 0]$ has the same support as $f$.  By
    \cite[Lemma~11]{DGJ:weighted-Boolean}, $\numCSP(f^2)$ is
    \numP-hard and $\numCSP(f^2) \reducesto \numCSP(f)$ by
    Lemma~\ref{lemma:squaring}.
\end{proof}

\begin{lemma}
\label{lemma:not-product-type-with-theta}
    If $f\in\ratfunc$ is not of product type of degree at most one
    then the problem $\numCSP(\{f, \delta_0, \delta_1, \Theta_\lambda\})$ is
    \numP-hard for any positive rational $\lambda\neq 1$.
\end{lemma}
\begin{proof}
    If $f$ is not of product type then, by
    Lemma~\ref{lemma:square-preserve}, $f^2\in\ratfunc[\geq 0]$ is
    also not of product type.  By
    \cite[Lemma~15]{DGJ:weighted-Boolean}, $\numCSP(\{f^2\!, \delta_0,
    \delta_1, \Theta_\lambda\})$ is \numP-hard for any positive,
    rational $\lambda\neq 1$ and the result follows by
    Lemma~\ref{lemma:squaring}.

    If $f$ is of product type but of degree two or more, the result
    follows from Lemma~\ref{lemma:product-type-high-degree}.
\end{proof}

The next lemma corresponds to \cite[Lemma~16]{DGJ:weighted-Boolean}
and its proof is based on the same idea as the proof there.  The
only difference is a slight adjustment to deal with mixed signs.

\begin{lemma}
\label{lemma:hardness-with-pinning}
    If $f\in\ratfunc$ is not pure affine of degree at most two and
    $g\in\ratfunc$ is not of product type of degree at most one, then
    $\numCSP(\{f, g, \delta_0, \delta_1\})$ is \numP-hard.
\end{lemma}
\begin{proof}
    Suppose $f$ is not pure affine of degree at most two.  If $f$ does
    not even have affine support, we are done by
    Lemma~\ref{lemma:not-affine} and, if $f$ is pure affine of degree
    three or higher, we are done by
    Lemma~\ref{lemma:pure-affine-high-degree}.  So we may assume that
    $f$ is not pure affine.  By Lemma~\ref{lemma:square-preserve},
    $f^2$ is also not pure affine and, by Lemma~\ref{lemma:squaring},
    it suffices to show that $\numCSP(\{f^2\!, g, \delta_0,
    \delta_1\})$ is \numP-hard.

    Since $f^2$ has affine support but is not pure affine, there must
    be at least two positive values in its range.  The proof now
    proceeds exactly as that of Lemma~16 in
    \cite{DGJ:weighted-Boolean}.  By using pinning and projection, we
    extract from $f^2$ a unary function $\Theta_\lambda$ for some
    positive rational $\lambda \neq 1$.  The function $\Theta_\lambda$
    is simulated by $f^2$ and we show hardness of $\numCSP(\{f^2\!, g,
    \delta_0, \delta_1\})$ by reduction from $\numCSP(\{g, \delta_0,
    \delta_1, \Theta_\lambda\})$, which is \numP-hard by
    Lemma~\ref{lemma:not-product-type-with-theta}.  We do not repeat
    the details here; refer to the proof in
    \cite{DGJ:weighted-Boolean}, starting with the second paragraph
    and noting that the function $g$ referred to there is the function
    $f^2$ here.
\end{proof}


\bibliographystyle{plain}
\bibliography{mixed_signs}

\end{document}